\documentstyle[12pt,epsfig]{article}
\newcommand{\PL}[3]{{Phys. Lett.}        {\bf #1} {(19#2)} {#3}}
\newcommand{\PRL}[3]{{Phys. Rev. Lett.} {\bf #1} {(19#2)} {#3}}
\newcommand{\PR}[3]{{Phys. Rev.}        {\bf #1} {(19#2)} {#3}}
\newcommand{\NP}[3]{{Nucl. Phys.}        {\bf #1} {(19#2)} {#3}}
\newcommand{\PRe}[3]{{Phys. Rep.}        {\bf #1} {(19#2)} {#3}}
\newcommand{\PTP}[3]{{Prog. Theor. Phys.} {\bf #1} {(19#2)} {#3}}
\newcommand{\ZP}[3]{{Z. Phys.}        {\bf #1} {(19#2)} {#3}}
\newcommand{\MPL}[3]{{Mod. Phys. Lett.}        {\bf #1} {(19#2)} {#3}}
\newcommand{\beq}{\begin{equation}}
\newcommand{\eeq}{\end{equation}}
\newcommand{\ba}{\begin{array}}
\newcommand{\ea}{\end{array}}
\newcommand{\beqa}{\begin{eqnarray}}
\newcommand{\eeqa}{\end{eqnarray}}
\newcommand{\dis}{\displaystyle}
\newcommand{\gsim}{\stackrel{>}{_\sim}}
\newcommand{\lsim}{\stackrel{<}{_\sim}}
\newcommand{\no}{\nonumber}
\newcommand{\da}{\dagger}
\newcommand{\Real}{\Re e}
\newcommand{\wt}{\widetilde}
\newcommand{\Lto}{\stackrel{\mu\to \Lambda_C}{\longrightarrow}}     
\renewenvironment{thebibliography}[1]
	{\frenchspacing
	 \small
	 \baselineskip=11pt
	 \begin{list}{[\arabic{enumi}]}
	{\usecounter{enumi}\setlength{\parsep}{0pt}
	 \setlength{\leftmargin 12.7pt}{\rightmargin 0pt}
	 \setlength{\itemsep}{0pt} \settowidth
	{\labelwidth}{#1.}\sloppy}}{\end{list}}

\begin{document}

\begin{flushright}
INPP-UVA-96/04 \\
LNF/96-051(P)  \\
hep-ph/9609518
\end{flushright}
\begin{center}

\vspace*{0.7 cm} 
{\large \bf Anatomy of the Higgs mass spectrum}  
\vspace*{1.0 cm}

{\bf{P.Q. Hung$^1$ and G. Isidori$^2$}}
\vspace{0.5cm}

${}^{1)}$ Physics Department, University 
of Virginia, Charlottesville, VA 22901, USA

${}^{2)}$ INFN, Laboratori Nazionali di Frascati,
P.O. Box 13, I--00044 Frascati, Italy

\end{center}
\vspace{0.5 cm}

\begin{abstract}    
\noindent
We analyze the implications of a Higgs discovery 
on possible ``new--physics''  scenarios, 
for $m_H$ up to  $\sim 700$ GeV. 
For this purpose we critically review 
lower and upper limits on the Higgs mass in the SM
and in the MSSM, respectively. Furthermore, we discuss the 
general features of possible ``heavy'' ($m_H \gsim 2 m_Z$)  Higgs 
scenarios by means of a simple heavy--fermion condensate model. 
\end{abstract}

\vglue 1.0 true cm

\section{Introduction}

The discovery of the Higgs particle is of utmost importance in particle
physics. Over the years, various theoretical bounds have been made
\cite{Cabibbo,Lindner,SherRep,HunterGuide,Schrempp,Sher,AI,CEQ1},  
and most recently 
an experimental {\em lower} bound of 65 GeV was set \cite{LowerB}.  
But the Higgs boson still remains elusive.  
Its nature --mass and couplings-- would reveal
the most fundamental aspects of the kind of mechanism that governs the
spontaneous symmetry breakdown of the Standard Model (SM). In
particular, one would like to know whether or not such a discovery, if
and when it will be made, will be accompanied by ``new physics'' at some
energy scale $\Lambda$. Of equal importance is the following question:
at roughly what mass scale will the Higgs boson be considered
elementary or composite? Can one make some meaningful statement
concerning its nature once it is discovered? These are the issues we
would like to explore in this paper. 

A first step in this direction has been recently achieved by
detailed analyses of the Higgs potential \cite{Sher,AI,CEQ1}. Indeed,
with the discovery of the top quark 
with mass $m_t = 175 \pm 9$ GeV \cite{CDF}, the Higgs mass ($m_H$)
is severely constrained by the requirement of vacuum stability. In particular,
two interesting conclusions have been drawn:
\begin{enumerate}
\item[{\bf i.}] If a Higgs will be discovered at LEP200, i.e. with 
with $m_H\leq m_Z$, then some new physics must appear
at very low scales: $\Lambda \lsim 10$ TeV \cite{AI,CEQ1,Hung}.
\item[{\bf ii.}] The Standard Model with an high cut--off 
(without new particles below $10^{15}$ GeV) 
requires $m_H \gsim 130$ GeV and is 
incompatible with the  Minimal Supersymmetric Standard Model (MSSM), where 
the mass of the lightest Higgs boson is expected below $130$ GeV \cite{CEQ1}.
\end{enumerate}

We shall re--analyze the above statements,
trying to clarify the stability of the physical 
conclusions with respect to the theoretical errors, and
we shall extend the discussion studying the implications
of a Higgs discovery up to to approximately 700 GeV.

The plan of the paper is as follows.
In section II and III we shall review what is
known about the Higgs sector of the SM and of the MSSM. 
We then divide our analysis into three 
separate mass regions: the region below $m_Z$,
between $m_Z$ and $2m_Z$, above $2m_Z$.

\section{The Higgs boson in the minimal Standard Model}

The symmetry breaking sector of the SM is highly unstable and 
make sense only in presence of a cut--off scale $\Lambda$. The 
instability of the scalar sector implies that upper and lower limits 
on $m_H$, imposed by the requirement of no Landau pole and 
vacuum stability, both below $\Lambda$, tend to shrink together as the 
cut--off increases \cite{Cabibbo,Lindner}.  

The instability of the scalar potential is generated by the quantum loop 
corrections to the classical expression 
\beq
V^{tree}(\Phi) =-m^2\Phi^\da\Phi+{\lambda \over 6}(\Phi^\da\Phi)^2,
\qquad\qquad \Phi=\left(\ba{c} \phi^+ \\  \phi^0 \ea \right)
\label{treepot1}
\eeq
where $v^2 = 2 \langle \phi^0 \rangle^2 =1/(\sqrt{2}G_F)\simeq 
(246~\mbox{GeV})^2$ and $\phi = (\sqrt{2} \Real \phi^0 -v)$ is the 
physical Higgs field. As already noticed in Ref.~\cite{Cabibbo}, and 
successively confirmed by 
detailed analysis of the renormalization group (RG) improved  potential  
\cite{SherRep,Ford}, the issue of vacuum stability for 
$\phi\sim\Lambda \gg m_Z$ practically coincides 
with the requirement that the running coupling $\lambda(\Lambda)$
never becomes negative. On the other hand, the requirement that no Landau
pole appears before $\Lambda$ is equivalent to the condition 
that $\lambda(\Lambda)$ always remains in the perturbative region.

The evolution of 
$\lambda$ as a function of $\Lambda$ is ruled by a set of coupled 
differential equations
\beqa                           
\ba{ccc}                      
\mbox{\rm d}\lambda(t)/\mbox{\rm d}t &=&\beta_\lambda(\lambda,g_i),  \\
\mbox{\rm d}g^2_i(t)/\mbox{\rm d}t &=& \beta_i(\lambda,g_i), \ea
\qquad\qquad t=\ln\left(\Lambda/\mu\right)
\label{evol1} 
\eeqa
with the corresponding set of initial conditions which relate 
$\lambda(\mu)$ and $g^2_i(\mu)$ to physical observables 
($g_3$, $g_2$ and $g_1$ denote   
$SU(3)_C\times SU(2)_L\times U(1)_Y$ gauge couplings and 
$g_{t}$ the top--quark Yukawa coupling, all couplings 
are understood in the $\overline{MS}$ scheme).
The $\beta$--functions of Eq.~(\ref{evol1}) are known in perturbation theory 
up to two loops (see Ref.~\cite{Ford,Schrempp} for 
the complete expressions), i.e. up to the third order in the expansion 
around zero in terms of $\lambda$ and $g_i^2$, 
whereas the finite parts of the initial 
conditions around $\mu=m_Z$ (threshold corrections) are 
known up to one--loop accuracy 
\cite{Sirlin,Arason,Hempfling}. This knowledge enable us to re--sum 
all the  next--to--leading logs in the evolution of the coupling constants
and thus to calculate them with high accuracy in the perturbative region. 
Nevertheless, the instable character of $\lambda(t)$  
can be simply read--off  by the one--loop expression
\beq
\beta_\lambda ={1\over 16 \pi^2} \left[ 4\lambda^2 +12 \lambda g_t^2 -36 g_t^4 +
O(g_1^2,g_2^2) \right],
\label{lambda1}
\eeq
together with the tree--level relations
\beq
\lambda(m_H) = {3 m_H^2 \over v^2 } \qquad\quad \mbox{and} \qquad\quad 
g_t^2(m_t) = {2 m_t^2 \over v^2 }.
\label{thres0}        
\eeq
For small values of $m_H$ the $g_t^4$ term 
in Eq.~(\ref{lambda1}) drives $\lambda$ to negative values, 
whereas  if $m_H$ is large enough the Higgs 
self--interaction dominates and eventually $\lambda$ ``blow--up''.

The situation is summarized 
in fig.~1 where we plot the evolution of $\lambda$
as obtained by integrating two--loop beta functions.
For $m_t=175$ (pole mass) and $\alpha_S(m_Z)=0.118$, if we impose the condition
\beq
0 < \lambda(\Lambda) <10 
\label{cond1}
\eeq
at the Planck scale, then $m_H$ is confined in a very narrow range
(full lines in fig.~1):\footnote{~With respect to Ref.~\cite{AI} we have 
removed a small error in the threshold correction of $g_t$ (the 
correct expression is given in Ref.~\cite{Hempfling}) 
obtaining a $\sim 1$ GeV decrease of the lower limit.}  
\beq
136~\mbox{GeV} \leq m_H \leq 174~\mbox{GeV}.\label{stablerange}
\eeq 
The lower limit on $m_H$ depends strongly on the values of $m_t$ and 
$\alpha_S(m_Z)$ \cite{Sher,AI,CEQ1} whereas the upper one is more or 
less independent from them, both are weakened if the condition 
(\ref{cond1}) is imposed at scales $\Lambda$ below the Planck mass 
(dashed lines in fig.~1). What happens if $m_H$ is outside the 
range (\ref{stablerange})?

For what concerns the problem of vacuum instability, the usual wisdom 
asserted that new physics must show up at or before the scale $\Lambda$
where $\lambda(\Lambda)$ becomes negative. However, as shown 
recently in Ref.~\cite{Hung}, the physical meaning of the previous
statement is not trivial. In particular, there are models 
where the masses of the new particles could be substantially larger 
than $\Lambda$ and still stabilize the vacuum.
We shall come back to this in section IV.

The upper bound in Eq.~(\ref{cond1}) can be considered as an 
upper limit for the applicability of perturbation theory ($\lambda/4\pi$
is the expansion parameter) and indeed below this value the difference between 
one-- and two--loop beta functions is not large (dotted curve in fig.~1).
However, for $m_H\simeq 180$ GeV, i.e. just above 
the upper limit imposed by Eq.~(\ref{cond1}),
the integration of one--loop beta functions originates a singularity at
$\Lambda_L < M_{Planck}$. As the Higgs mass increases 
$\Lambda_L$ decreases and approaches $10^{5}$ GeV for
$m_H \approx$ 300 GeV. What is the physical meaning of the
singularity scale $\Lambda_L$? 
If one believes that the Landau pole is not an 
artifact of perturbation theory but a non--perturbative feature of the 
model, as suggested by lattice simulations (see e.g. 
Refs.~\cite{lattice,Schrempp}), then is tempting to 
think that some new physics must occur around $\Lambda_L$. 
If that is so, this
kind of new physics must be {\em very} different from the one
needed to stabilize the vacuum, since one is now dealing with a strong
coupling domain. One is then tempted to attribute this behaviour
(strong coupling) to the nature of the Higgs boson. In particular,
one might think that the Higgs boson is a composite particle which
acts like an elementary field below the scale $\Lambda_L$. How one
can tell if this is the case is the subject of our discussion in
section VI.

\section{The Higgs sector of the MSSM}

The Higgs sector of the MSSM (see Refs.~\cite{Susy} for excellent reviews)
contains two Higgs doublets, one is responsible for charged--lepton and 
down--type--quark masses ($H_1$), the other  for
up--type--quark masses ($H_2$). Of the eight degrees of freedom, two charged, 
one $CP$--odd and two $CP$--even neutral scalars
correspond to physical particles after the $SU(2)_L\times U(1)_Y$ 
breaking. The tree--level potential for the two neutral components 
$H_1^0$ and $H_2^0$ is given by
\beq
V^{tree}(H^0_1,H^0_2) = { g_1^2 + g_2^2 \over 8}  (|H_1^0|^2 - |H_2^0|^2)^2
+ m^2_1 |H_1^0|^2 + m^2_2 |H_2^0|^2 + [m^2_{12} H_1^0H_2^0 + \mbox{h.c.}].
\label{treepot2}
\eeq
The sum of the two vacuum expectation values squared 
is fixed by the gauge boson masses: $v^2_1+v^2_2=v^2$ 
($v_{1,2}=\sqrt{2}\langle H^0_{1,2} \rangle$), while  
the ratio $\tan(\beta)=v_2/v_1$ is a free parameter. The remarkable feature 
of the potential~(\ref{treepot2}) is that the coefficient of the 
dimension four operator
is completely fixed in terms of the gauge couplings $g_1$ and $g_2$. 
This property leads to the tree--level relation  
\beq
m^2_{H,H'} = {1\over 2}\left( m_A^2 + m^2_Z \mp \sqrt{ ( m_A^2 +
m_Z^2 )^2   -4m_A^2 m_Z^2 \cos^2 2\beta } \right),
\eeq
where $m_{H,H'}$ are the two $CP$--even Higgs boson masses
and $m_A$ is the $CP$--odd one, which implies a strict upper bound  
\beq
m_H \leq m_Z \cos 2\beta \leq m_Z
\label{treeboud}
\eeq  
on the lightest Higgs boson mass. 

As it is well--known~\cite{SusyLim,SusyLim2,CEQ1}, 
the bound~(\ref{treeboud}) receive large radiative corrections 
if SUSY particles, and in particular the ${\wt t}$ squark, are heavy. 
This can be easily understood by means of the SM evolution of $\lambda$
previously discussed. Indeed, if all SUSY particles (including 
additional Higgs bosons) have a mass of the order of $M_S$ 
($M_S^2  \gg m^2_Z$), the lightest Higgs boson decouples 
below $M_S$ and mimics the SM Higgs. Then, the evolution of the scalar
self-coupling $\lambda(\Lambda)$ is dictated by SM beta functions up to 
$\Lambda=M_S$, 
where SUSY is restored and, according to the potential~(\ref{treepot2}),
the following relation must old:
\beq
\lambda(M_S)={3\over 4}\left[ g_1(M_S)^2 +g_2(M_S)^2 \right]\cos^2 2\beta.
\label{Susycond1}    
\eeq
Eq.~(\ref{Susycond1}) saturates the bound~(\ref{treeboud})
for $M_S\sim m_Z$ but, due to the rapidly decreasing behaviour
of $\lambda(\Lambda)$ (see fig.~2), implies $(20\div 30)\%$ violations 
of the tree--level bound for $M_S\sim 1$ TeV \cite{SusyLim}.

Analogously to the tree--level relations (\ref{thres0}), the boundary 
condition~(\ref{Susycond1}) is not differentiable with respect to the 
scale of $\lambda$: in order to calculate precise 
bounds on the Higgs mass is necessary to
include threshold effects in both cases. The most important correction to
Eq.~(\ref{Susycond1}) is the one generated by stop loops, that is 
proportional to $g_t^4$. If we include this effect 
Eq.~(\ref{Susycond1}) is modified in 
\beq
\lambda(\Lambda)={3\over 4}\left[ g_1(\Lambda)^2 +g_2(\Lambda)^2 \right]
\cos^2 2\beta +\Delta \lambda(\Lambda),
\label{Susycond2}     
\eeq
where
\beq
{\mbox{d} \Delta \lambda(\Lambda) \over \mbox{d}\ln (\Lambda/\mu)} =
-{36 \over 16 \pi^2} g_t^4 +...,
\label{delta00}    
\eeq
by this way the leading term in the derivative of both sides of 
Eq.~(\ref{Susycond2}) is the same. The explicit expression  of 
$\Delta \lambda(\Lambda)$, obtained by the one--loop 
stop contribution to the potential~(\ref{treepot2}), is given by
\cite{CEQ1}
\beqa
\Delta \lambda(\Lambda) = &\dis{9 g_t^4 \over 16 \pi^2}& \left\{ 
{ (m_t+X_t)^2 \over  {\wt m_+}^2 } \left[ 1 -
{ (m_t+X_t)^2 \over  12{\wt m_+}^2 } \right] \right. \no \\
&& \left. +  { (m_t-X_t)^2 \over  {\wt m_-}^2 } \left[ 1 -
{ (m_t-X_t)^2 \over  12{\wt m_-}^2 } \right] +\ln\left(
{{\wt m_-}^2 {\wt m_+}^2  \over \Lambda^4} \right) \right\},
\label{delta01}    
\eeqa
where ${\wt m_\pm} = M_S^2 +m_t \pm m_tX_t$ are the eigenvalues of the 
stop mass matrix and $X_t$ is the usual stop--mixing
parameter \cite{CEQ1,SusyLim}. As noticed in 
Ref.~\cite{CEQ1}, $\Delta \lambda(M_S)$ has a maximum for 
$X_t^2= 6M_S^2 +O(m_t^2)$.

Imposing the boundary condition (\ref{delta01}) at $\Lambda\sim M_S$,
using two--loop SM beta functions to evolve down at $\mu\sim m_Z$ 
and finally using SM one--loop matching conditions to relate 
$m_H$ and $m_t$ to $\lambda$  and $g_t$, we find (masses are in units of
GeV):
\beqa
M_H^{MSSM} < 127  
+0.9\left[m_t-175\right]
-0.8\left[\dis{\alpha_S(m_Z)-.118 \over .006} \right] +7\cdot\log_{10}
\left( {M_S \over 10^3 }\right) \pm 4 
\label{Susy127}  
\eeqa
in good agreement with the detailed analysis of Ref.~\cite{CEQ1}.
The error in Eq.~(\ref{Susy127}) has been estimated by varying low and high 
energy matching scales in the following intervals: 
$\Lambda \in [M_S, 2M_S]$ and $\mu \in [m_Z, 2 m_t]$.
Obviously the upper limit is very sensitive to $M_S$, defined as the soft stop
mass, and is valid for $M_S$ near $1$~TeV; on the other hand, 
the dependence form other SUSY masses is within the quoted error.  
As can be noticed from fig.~2, for $m_t=175$~GeV and 
$\alpha_S=0.118$, the SM with $\Lambda=M_{Plank}$ is compatible 
with the MSSM only for unnatural large values of $M_S$.

\section{Physics of the ``low'' mass Higgs boson: $m_H \leq m_Z$ GeV}

As we have discussed in section II, the SM becomes unstable when
$m_H \leq 136$ GeV. Moreover, if the Higgs mass is below the $Z$ mass, 
the SM breaks down at a scale $\Lambda$ situated in the TeV region
\cite{AI,CEQ1,Hung}.

Recently, it has been pointed out \cite{CEQ2} that for small values
of $\Lambda$ the lower limit on $m_H$ imposed by the condition 
\beq
\left. {\mbox{d} V^{1-RG}(\phi) \over \mbox{d} \phi}\right\vert_{\phi=\Lambda}
>0, \label{cond2}
\eeq
where $V^{1-RG}(\phi)$ denotes the one--loop  RG--improved potential,
do not coincide with the one imposed by $\lambda(\Lambda)>0$. We agree 
with the above statement, 
however it must be stressed that the two conditions 
lead to equivalent results up to a
small re--definition of $\Lambda$ \cite{CEQ2}. As an example,
the lower limit on $m_H$ imposed by Eq.~(\ref{cond1}) with 
$\Lambda=1$~TeV, namely  $m_H> 72$ GeV,  is equivalent to the one imposed  
 by Eq.~(\ref{cond1}) with $\Lambda \simeq 3.4$~TeV. On the other hand, 
the two conditions coincide for large values of the cut--off, where the 
corresponding $\lambda(\Lambda)$ curves are almost flat (fig.~1).
Since the exact relation between $\Lambda$, understood as the scale where the 
evolution of $\lambda$ is no more ruled by  Standard Model beta functions, 
and the masses of 
hypothetical new particles depends on the details of the new--physics
model \cite{Hung},
in our opinion is meaningless to fix $\Lambda$
with great accuracy. In other words, for a given value of $m_H$, the scale 
$\Lambda$ where Eq.~(\ref{cond1}) or Eq.~(\ref{cond2}) are no more 
satisfied can give only an indication of the order of magnitude below
which new physics must appear, and within this interpretation the two
conditions are completely equivalent and consistent with the 
statement {\bf i} of sect.~I.                   

To stabilize the SM vacuum, one has to add more scalar degrees of
freedom which couple to the SM Higgs, a well-known fact 
from studies of the effective potential
or from studies of the RG equations.
The most natural new--physics candidate in this case 
is the MSSM.  There there is
a plethora of scalar fields: the supersymmetric partners of 
quarks and leptons, and the additional Higgses. However, 
as we have seen in the previous section, the ``stabilizing scalar''
is the stop which cancel the $g_t$ dependence in the evolution of
$\lambda$. More light is the Higgs and more light must be the stop.   

What happens if the Higgs mass is very light,
say 70 GeV, and the top is not found in the TeV region? 
It could mean several things. Either the MSSM is not correct and
a more complicated version is needed or something other than SUSY
enters the picture. In Ref.~\cite{Hung} this question has been 
studied using a toy model with electroweak singlet scalars, 
with multiplicity $N$ and with a coupling $\delta$ to the standard
Higgs field. It was found that the mass of the new singlet scalars
could be as high as ten times the scale $\Lambda$ where $\lambda(\Lambda)$
becomes negative.

In the above discussion, there was never any need for the
Higgs boson to be composite. In fact, it appears to be more natural
for the Higgs boson to be {\em elementary} in this case. Although
there are models for a ``light'' Higgs boson where an elementary Higgs
field is mixed with a top condensate \cite{Gerard}, it does not appear
to be possible to construct a model where the Higgs boson is
entirely composite. It is in this sense that we say that the Higgs
boson is elementary if its mass is $m_Z$ or below.

The main conclusion of this section is the following: if $m_H
\leq m_Z$, the Higgs boson is most likely elementary 
and there should be new physics, within the 10 TeV scale,
either in terms of SUSY particles or in terms of new scalar degrees
of freedom.

\section{Physics of the Higgs boson with $m_Z \leq m_H \leq 2 m_Z$}

This is a region where it will be extremely hard to detect the
Higgs boson \cite{HunterGuide}. 
Theoretically, this is a region where one can still 
presume that the Higgs boson is an elementary particle. Indeed, a
look at fig.~1 will convince us that $\lambda$ blows up below the
Planck scale only when $m_H \gsim 2 m_Z$. Furthermore, there is no known
mechanism which can give rise to a composite Higgs boson that light
(without additional scalars).

As we have seen in section III, if $m_H \lsim 130$ GeV the most natural 
candidate is still the MSSM. On the other hand, 
for $m_H \gsim 130$ GeV the MSSM it is unnatural because 
the SUSY scale is too high.  Above 130 GeV natural candidates are
SUSY extensions of the SM with a non--minimal scalar sector 
\cite{Espinosa2}. In this region also the SM itself can be considered
a good candidate. Indeed, a part from the problem of 
quadratic divergences, new--physics can be pushed up to the Planck scale
if $m_H \gsim 130$. In this framework, an interesting scenario 
is the one proposed in Ref.~\cite{FN}.  

\section{Physics of the Higgs boson with $m_H \geq 2 m_Z$}

We finally come to the  
question of which kind of new physics is 
expected if $m_H$ if found above $\sim 180$ GeV, i.e.
in the region where $\lambda(\Lambda)$ develops a singularity at 
$\Lambda_L < M_{Planck}$. As we have already said in sect. I, 
the Landau pole might just be an
artifact of perturbation theory. However we believe this is not 
the case. Following the indications of   
lattice simulations \cite{lattice}, we believe that the 
presence of such singularity is at least qualitatively
correct and that indicates the {\em composite nature} 
of the Higgs boson.  

The physics below the compositeness scale can be described in terms
an effective field theory whose couplings are constrained by the
boundary conditions at the compositeness scale. In this framework, 
a class of models which is particularly attractive, relevant
to the present discussion and quite general is the  
class of the top--condensate models \cite{Topcond1,BHL}. 
There the relevant boundary conditions are \cite{BHL,BLP}:
\beqa
&& \lambda(\mu),~g_t(\mu)   \Lto  \infty  \label{boundc1} \\
&& \lambda(\mu)/g^2_t(\mu)  \Lto  \mbox{const.}
\label{boundc2}
\eeqa          
Thus the Landau singularity of the 
Higgs self-couplings naturally fits into this scheme.
The only problem is the requirement of a  pole
also in the evolution of the top Yukawa coupling. 
As can be noticed in fig.~3, the top Yukawa coupling itself is not large 
enough to develop a singularity since its evolution is  ``softened'' by QCD.
However, as we will show in the following,
if we include additional heavy fermions with a mass 
$m_f$ above a critical value, both $g_t$ and $g_f$ can ``blow up''
at a scale $\Lambda_f$.

To analyze better the model, let us consider the Lagrangian
of a single degenerate quark doublet
$q = (u, d)$ coupled to the Higgs field. 
If we re--scale the Higgs field in the following way
\beq
\Phi \longrightarrow \Phi_{0} / g_f,
\eeq
the Lagrangian becomes
\beq
{\cal L}= {\cal L}_{kinetic}(u,d) +
 Z_{\Phi} D_{\mu}\Phi^{\dagger}_{0} D^{\mu}\Phi_{0} 
+{\widetilde{m}}^2
\Phi^{\dagger}_{0} \Phi_{0} -\frac{\widetilde{\lambda}}{6}
(\Phi^{\dagger}_{0} \Phi_{0})^2
+ {\bar q}_{L}\Phi_{0} d_R +{\bar q}_{L}{\Phi^C_{0}} u_R + \mbox{h.c.},
\eeq
where
\beq
\Phi^C_0 = i\sigma_2 \Phi_0^{\ast} , \qquad
Z_{\Phi} = 1/ g^2_f,  \qquad
\widetilde{m}^2 = Z_{\Phi} m^2 , \qquad\mbox{and}\qquad  
\widetilde{\lambda} = Z_{\Phi}^2\lambda.
\eeq
If $\lambda$ and $g^2_f$ develop a singularity 
at the same scale $\Lambda_f = \Lambda_L =\Lambda_C$, so that the
boundary conditions (\ref{boundc1}-\ref{boundc2}) are satisfied, then
with an appropriate tuning of the  quadratic divergences
we can have \cite{BLP}
\beq
\widetilde{m}^2/\Lambda_C^2   \Lto  \mbox{const}  <0.
\eeq
Thus at the compositeness scale the above Lagrangian becomes
\beq
{\cal L}= {\cal L}_{kinetic}(u,d) + 
{\bar q}_{L}\Phi_{0} d_R +{\bar q}_{L} \Phi^C_{0} u_R 
+{\widetilde{m}}^2 \Phi^{\dagger}_{0} \Phi_{0} +\mbox{h.c.} 
\eeq
and $\Phi_0$, which is now just an auxiliary field, can 
can be integrated out to obtain 
a four--fermion Nambu--Jona--Lasinio Lagrangian \cite{NambuY}
\beq
{\cal L} = {\cal L}_{kinetic}(u,d) + G_0 {\bar q}_L( u_R {\bar u}_R + 
d_R {\bar d}_R) q_L ,
\eeq
with $G_0 = -1/\widetilde{m}^2.$
Viewed in this way, the Higgs boson becomes 
a dynamical fermion--condensate below the
scale $\Lambda_C$, in other words 
the Higgs boson becomes a composite particle.

The necessary conditions for the above view to 
hold are the constraints (\ref{boundc1}-\ref{boundc2}).
In order to understand if these conditions can be 
satisfied, it is useful to examine the 
RG equation of the ratio $x=\lambda/g^2_f$.
At one loop, it is given by
\beq
16 \pi^2 \frac{dx}{dt} = 4 g^2_f (x - x_{+})(x - x_{-}) ,
\eeq
where $x_{\pm} = (-3 \pm 9)/2$,
if both members of the quark doublet are degenerate in mass, or
$x_{\pm} = \frac{3}{8} (-1 \pm \sqrt{65})$,
if one member is much heavier than the other one 
(e.g. the 3rd generation case). 
The only possibility to have $\Lambda_f = \Lambda_L $  
is that the initial value of $x$  is one of the two fixed points.
Since $x_{-}$ is always negative, the solution 
$x= x_{-}$ is ruled out by vacuum stability. Thus
the boundary conditions can be satisfied
only if  $x=x_{+}$ and this implies a precise relation between 
$m_H$ and $m_f$. Using the tree--level relations (\ref{thres0}) we find 
\beq
m^2_H = \frac{2}{3} m^2_f x_{+}.\label{fixedp}
\eeq   
Since $x_{+}$ is always greater 
than $3/2$, the fixed point scenario implies
\beq
m_H > m_f. \label{mhgtmf}
\eeq
It is easy to see that, in the large $N_c$ limit, the fixed point takes on the
value $x= 6$, giving $m_H = 2 m_f$,
a familiar result found in the Nambu-Jona-Lasinio model. 
For finite $N_c$ one finds in general 
\beq
m_f < m_H < 2 m_f.
\eeq

As we have mentioned earlier, the top quark is not heavy enough to 
solely fit into this scenario.
This is because the growth of the top Yukawa coupling is dampened by QCD. 
The minimum top mass for which there will be a 
Landau singularity at the Planck mass is $m_t \approx 216$ GeV, 
a value which is way outside the experimental range. 
Let us then assume that there is an extra doublet of
degenerate quarks, $Q= (U, D)$, whose mass is arbitrary. As
we shall see below, the addition of this extra doublet changes 
dramatically the behaviour of
the couplings at high energy. To see this let us write the RG 
equations for $\lambda$, $g_f$
(the new--doublet Yukawa coupling), and $g_t$ 
(the top Yukawa coupling):
\beqa
&& 16 \pi^2 \frac{d\lambda}{dt} = 4 \lambda^2 + 12 \lambda (g^2_t + 2 g^2_f) 
    - 36 (g^4_t + 2 g^4_f)  + O( g_1^2, g_2^2)  \label{gg1}    \\
&& 16 \pi^2 \frac{dg^2_f}{dt} = g^2_f [12 g^2_f + 6 g^2_t -16 g^2_3]
+ O( g_1^2, g_2^2)  \label{gg2}  \\
&& 16 \pi^2 \frac{dg^2_t}{dt} = g^2_t [9 g^2_t + 12 g^2_f -16 g^2_3 ]
+ O(g_1^2, g_2^2) .   \label{gg3}
\eeqa

In the absence of the extra quark doublet, one 
can easily see from the above equations
that the growth of $g_t$ is dampened by the gauge 
couplings (mainly by $g_3$)\footnote{~The $O( g_1^2, g_2^2)$ 
terms in Eqs.~(\ref{gg2}-\ref{gg3}), which tend to split the 
evolution of $U$ and $D$ Yukawa couplings, cannot be neglected
if we are interested in a precise determination of the critical value of
$g_f$ (see the discussion below). }.     
On the other hand, in presence of the extra doublet
$g_t$ is no longer dampened provided 
$g_f$ exceeds some critical value. In addition, $g_t$ and
$g_f$ tend to ``drag'' each other. If we allow the possibility
--not withstanding experimental constraints-- that there could
be an extra doublet of degenerate quarks with mass less 
than the top quark,\footnote{Note that electroweak 
precision data put severe constraints on possible
new--fermion mass splitting but there is still room 
for an additional degenerate fourth family 
of quarks and leptons \protect\cite{Langa}.}
then a numerical integration of the above equations 
shows that there is a minimum mass for the new fermions 
for which $g_t$ and $g_f$   
develop a singularity around the Planck scale. 
As shown in fig.~3, this minimum mass is $m_f\simeq 160$ GeV.
The corresponding Higgs mass,
determined by the condition that $\lambda$ develops a 
singularity at the same scale as $g_t$ and $g_f$,
is $m_H \simeq 190$ GeV. As $m_f$ increases, the 
compositeness scale $\Lambda_C$ decreases and 
the  relation between   $m_H$ and $m_f$ approaches the 
fixed point prediction (\ref{fixedp}) with $x_+ \simeq 3$ (see
fig.~3).

The above scenario cannot be considered as a realistic model. Indeed,
if the scale $\Lambda_C$ is high there is clearly a ``fine tuning''
problem related to the large disparity between $\Lambda_C$ and the
electroweak scale. However, it is beyond the scope of this paper to try 
to  construct an underlying theory around 
$\Lambda_C$ and thus we will ignore it.  
Our purpose is just to show  some general features of a wide 
class of models. In particular, if there are no new bosons 
(scalars or gauge bosons) below the compositeness scale, the following 
features hold independently of the multiplicity of the new fermions:
\begin{enumerate}
\item[{\bf i.}] The compositeness scale $\Lambda_C$, the heavy--fermion 
mass $m_f$, and the effective Higgs mass $m_H$, are tied 
together so that $m_H$ and $m_f$ increase as $\Lambda_C$ decrease.
\item[{\bf ii.}] As shown in Eq.~(\ref{mhgtmf}), one typically
finds $m_H > m_f$. Thus if $m_H$  is not found below $2m_Z$ it
should be ``easier'' to search for new fermions instead of searching 
for the Higgs boson itself. 
\item[{\bf iii.}] For $\Lambda_C \approx 1$ TeV both $m_H$ and $m_f$ 
are $O(\Lambda_C)$ and the Higgs effective theory 
becomes meaningless. In this sense we agree with the more precise 
and well--defined lattice bound: $m_H \lsim 700$ GeV \cite{lattice}.
\end{enumerate}

\section{Conclusions}
                               
In this paper we have analyzed the consequences of a Higgs
discovery up to approximately $700$ GeV, dividing the mass region 
into three parts: the region below $m_Z$,
between $m_Z$ and $2m_Z$, above $2m_Z$.
               
Regarding the first two regions
we have confirmed and refined the results
stated in the introduction, namely the SM lower bound due to 
vacuum stability and the MSSM upper bound. 

Regarding the last region ($m_H \gsim 2 m_Z$) we have shown,
by means of a simple heavy--fermion condensate model, how the 
Landau pole of the Higgs self coupling can be related to 
the compositeness of the Higgs particle. 
We have analyzed the  general features  
of such scenarios. In particular, we have shown
that there exists a precise relationship between 
the effective Higgs mass, 
the new--fermion mass and the compositeness scale, 
which should hold in a wide class of models. 

\bigskip

{\bf Acknowledgments}

\noindent
G.I. wishes to thank G. Altarelli and 
L. Maiani for interesting discussions, and the warm hospitality of the 
University of Virginia where part of this work was done. 
P.Q.H. wishes to thank the warm hospitality of the 
Theory Groups at the University of
Rome ``La Sapienza'' and at Fermilab where part of this work was done. 
P.Q.H. is supported in parts by the U.S. Department of Energy under 
grant No. DE-A505-89ER40518.

\newpage

\begin{figure}  
   \begin{center}
   \setlength{\unitlength}{1truecm}
       \begin{picture}(8.0,7.0)
       \epsfxsize   9.0 true  cm
       \epsfysize   9.0 true cm
       \epsffile{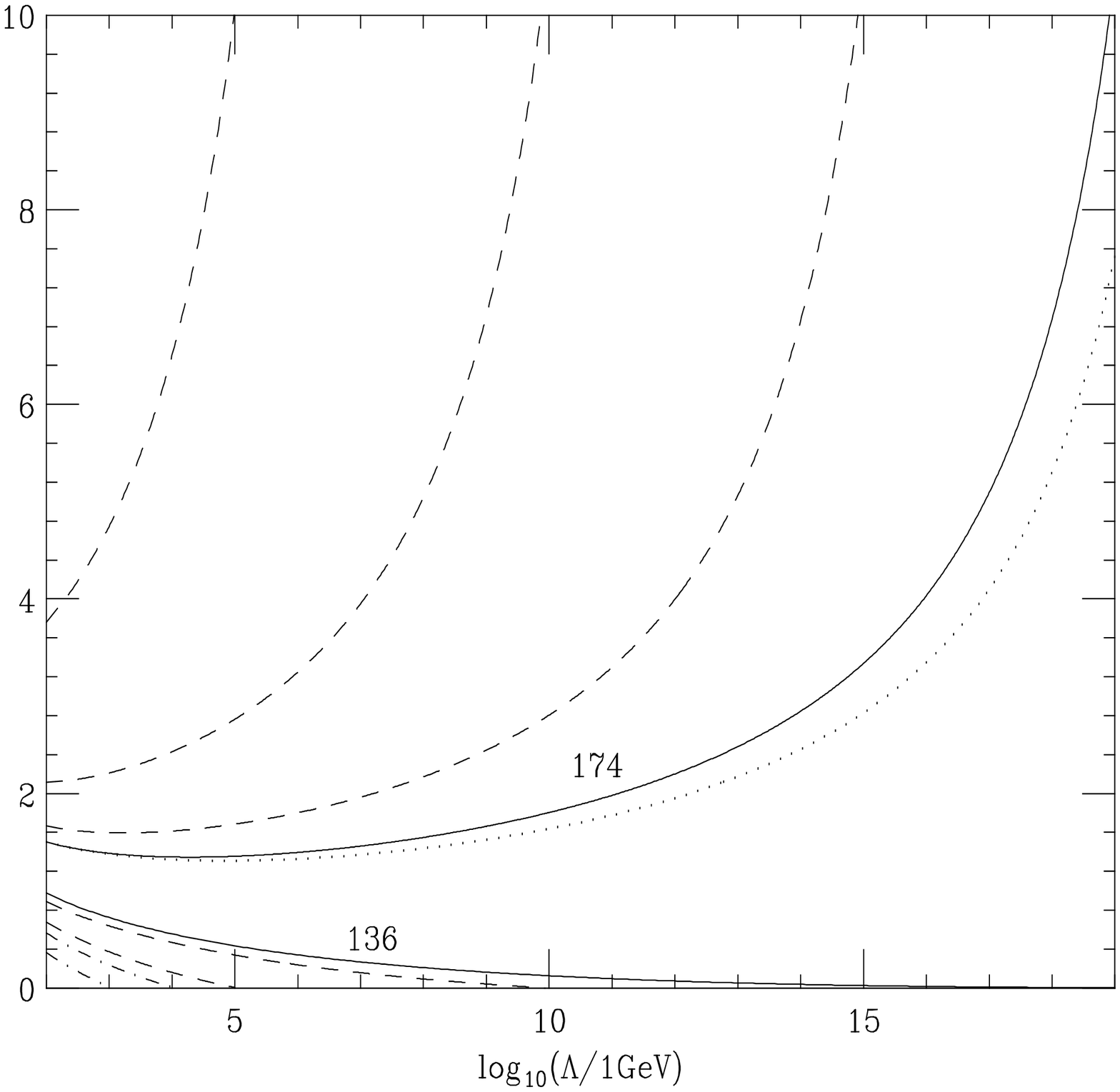}
   \end{picture}
   \end{center}
   \caption{\small $\lambda(\Lambda)$ for $m_t=175$ GeV (pole mass) and 
$\alpha_S(m_Z)=0.118$.  Full and dashed lines have been obtained 
by integrating 
two--loop beta functions and using  one--loop matching conditions,
whereas the dotted one has been obtained by integrating one--loop beta 
functions. The physical values of the Higgs mass 
corresponding to the full lines are $m_H=136$ GeV and $m_H=174$ GeV.  } 
   \protect\label{fig1}
\end{figure}

\begin{figure}  
   \begin{center}
   \setlength{\unitlength}{1truecm}
       \begin{picture}(8.0,7.0)
       \epsfxsize  9.0 true  cm
       \epsfysize  9.0 true cm
       \epsffile{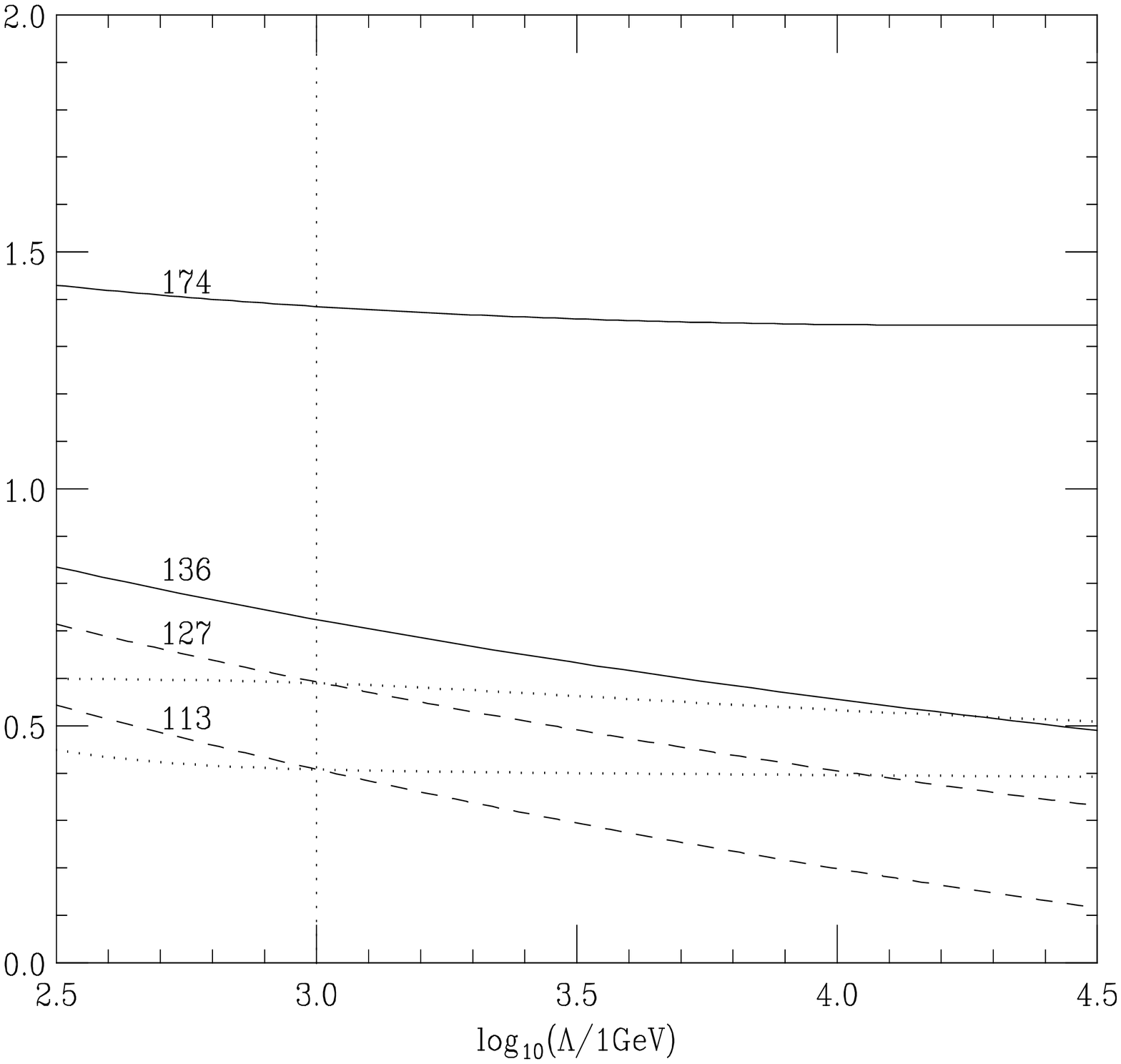}
   \end{picture}
   \end{center}
   \caption{\small 
$\lambda(\Lambda)$ in the small $\Lambda$ region. The horizontal
dotted lines indicate the MSSM upper limit imposed by 
Eq.~(\protect\ref{Susycond2}) for $M_S=\Lambda$ and $X_t=0$ (lower curve) 
or $X_t=X_t(M_S)$ chosen to maximize the threshold effect (upper curve). 
Full and dashed lines indicate the SM evolution of $\lambda$ 
(two--loop beta functions and one--loop matching conditions
with $m_t=175$ GeV and $\alpha_S(m_Z)=0.118$)
for different values of $m_H$ (as indicated above each line). }
   \protect\label{fig2}
\end{figure}

\begin{figure}  
   \begin{center}
   \setlength{\unitlength}{1truecm}
       \begin{picture}(8.0,7.0)
       \epsfxsize  9.0 true  cm
       \epsfysize  9.0 true cm
       \epsffile{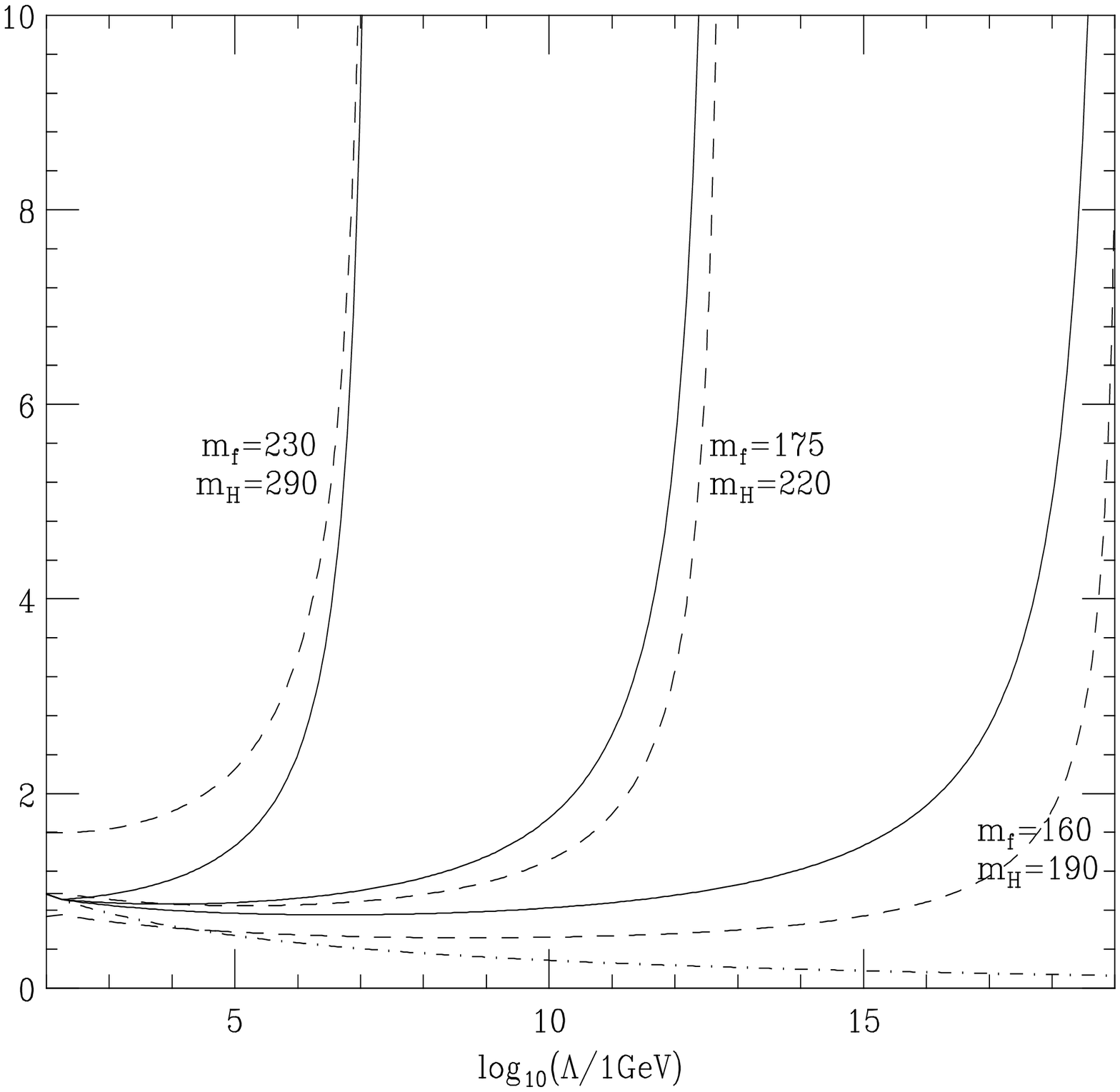}
   \end{picture}
   \end{center}
   \caption{\small
RG evolution of the Yukawa couplings $g_t^2$ (full and dash--dotted 
lines) and $g_f^2$ (dashed lines). The top mass is always fixed to be 175 GeV
and the dash--dotted line is the evolution of $g_t^2$ without the 
extra doublet. Near each dashed line is indicated the value of $m_f$ and 
the corresponding value of $m_H$ obtained by the requirement
$\Lambda_L=\Lambda_f$ (the error on both $m_H$ and $m_f$ is 
about $\pm 5$ GeV). }
   \protect\label{fig3}
\end{figure}

\end{document}